\documentstyle[preprint,aps]{revtex}

\def\arcsec{\hbox{$^{\prime\prime}$}}
\def\lesssim{\mathrel{\hbox{\rlap{\hbox{\lower4pt\hbox{$\sim$}}}\hbox{$<$}}}}
\def\gtrsim{\mathrel{\hbox{\rlap{\hbox{\lower4pt\hbox{$\sim$}}}\hbox{$>$}}}}
\def\arcsec{\hbox{$^{\prime\prime}$}}

\tightenlines

\begin{document}

\draft

\title{Gravitational Lensing Statistics in a Flat Universe}
\author{Myeong-Gu Park}
\address{Department of Astronomy and Atmospheric Sciences,\\
       Kyungpook National University, Taegu 702-701, Korea}
\maketitle

\medskip
\centerline{\it to appear in the Journal of Korean Physical Society}

\begin{abstract}

The probability distribution of lens image separations
is calculated 
for the ``standard'' gravitational lensing statistics model
in an arbitrary, flat Robertson-Walker 
universe, where lensing galaxies are 
singular isothermal spheres
that follow the Schechter luminosity function.
In a flat universe, the probability distribution 
is independent of the source
distribution in space and in brightness. 
The distribution is compared 
with observed multiple-image lens cases
through Monte-Carlo simulations and the Kolmogorov-Smirnov test.
The predicted distribution depends on the shape
of the angular selection bias used, which varies for different observations.
The test result also depends on which lens systems
are included in the samples. We mainly include the lens systems
where a single galaxy is responsible for lensing.
We find that the ``standard'' model predicts a distribution which is
different from the observed one. However, the statistical significance
of the discrepancy is not large enough to invalidate the ``standard'' model
with high confidence. 
Only the radio data reject the model at the 95\% confidence level,
which is based on four/three lens cases.
Therefore, we cannot say that the observational data
reject the ``standard'' model with enough statistical confidence.
However, if we take the velocity dispersion of dark matter
without the conversion factor $(3/2)^{1/2}$ from that of luminous matter,
the discrepancy is quite severe, and even
the ground-based optical survey data reject the ``standard'' model
with $\gtrsim 90\%$ confidence.

\end{abstract}

\pacs{PACS numbers: 95.30.S, 98.62.S, 98.54.Aj, 98.65.Dx, 98.80.Es}

\section{Introduction}\label{sec:intro}

Determining the geometry of the universe and the 
distributions of masses therein            
has always been one of the most important goals of astronomy.
Even before the actual discovery of gravitational lenses, 
it was known that lens systems could provide us much useful
information on these fundamental questions \cite{Ref64-Pre73}.

As more high-redshift multiply imaged QSOs were discovered, 
Turner et al. \cite{TOG} calculated
the statistical properties of multiple-image 
gravitational lenses in the standard Robertson-Walker cosmology,
especially the probability of a QSO being multiply imaged and
the mean angular separation of images \cite{TOG}. 
They compared the predicted
statistical properties of the lens systems with 
observations and concluded that
although more small-separation ($\lesssim 1 \arcsec$) lenses are
expected, the distribution of the image separation, with
the effect of resolution bias considered, is compatible with the
observed lens systems. Gott et al.
extended this work to an arbitrary Robertson-Walker universe \cite{GPL}. 
The probability of lensing and the mean image separation
were expressed as functions of the source redshift. 
However, the explicit 
value of the mean image separation was not calculated,
and only the functional form was compared with the data
to get the limits on the cosmological parameters. 
One interesting fact recognized in this work, and partly in 
Ref. \cite{TOG}, is
that in a flat universe filled with 
singular isothermal sphere (SIS) galaxies,
the mean separation of images is independent of the source
distance (or redshift),
regardless of the value of the dimensionless cosmological constant, 
$\lambda_0 \equiv \Lambda/3H_0^2$ where $H_0$ is
the Hubble constant.

Turner \cite{Tur80} and Fukugita et al. \cite{Fuk90} subsequently realized
that the probability of lensing, rather than the image separation, is
more sensitive to $\lambda_0$.
In a series of papers \cite{FT,FFKT}, they estimated how
many multiple-image lens systems should be discovered for given
QSO samples in a $\lambda_0$-dominated universe 
and found the present number of observed lens systems
exclude $\lambda_0 \gtrsim 0.9$. This limit
is considered as one of the few observational limits on
the elusive cosmological constant.

However, if
we naively use the various galaxy parameters of Ref. \cite{FFKT} to
recalculate the expected mean separation 
in a flat universe following Ref. \cite{GPL}, we get 
1.5\arcsec. This is significantly smaller than the
observed image separations in known single-galaxy lens systems
(systems in which a single galaxy is responsible for lensing), 
e.g., PG1115, Q2237, Q0142, H1413, MG0414, 
and B1422 (Table \ref{tableSAMPLE})
whose mean value is 1.95\arcsec and whose dispersion 0.70\arcsec.
If we
take the most favorable galaxy parameters allowed within
the uncertainty interval in Ref. \cite{FT}, the mean separation goes up to
1.7\arcsec, still somewhat small compared to the
observed image separations.
However, the angular resolution bias, larger
splitting lens systems being easier to find,
which was not considered in Ref. \cite{GPL} and which was
minimally incorporated in Refs. \cite{FT} and \cite{FFKT}
might change this interesting discrepancy.
Also, a better statistical test is needed
because of the non-standard shape of the distribution
and the small sample size.

The usual criticism against utilizing image separation data
has been its dependence on a combination of cosmological parameters
\cite{GPL}. However, this can be an advantage
if we want to test the curvature of the universe or the consistency 
of the lensing statistics model without 
being affected by the uncertainties in the
cosmological parameters. Moreover, the distribution of the image
separations is relatively independent of the magnification bias
which introduces major uncertainty when the probability of lensing
is used as a cosmological diagnostic tool.

In this work, we adopt the same assumptions and parameters
for galaxies as in Ref. \cite{FT} to get the probability
distribution of the image separations, incorporating
all possible angular resolution biases present in various lens surveys. 
By comparing this distribution with 
observed single-galaxy lens data (Table 1),
we test the ``standard'' gravitational lensing statistics model
against the observations.
We choose a flat universe with arbitrary $\lambda_0$ as the
background universe.

\section{Probability distribution of image separation}
\label{sec:PD}

A flat universe with a non-zero cosmological constant
can be described by the Robertson-Walker metric \cite{GPL}
\begin{equation}
   ds^2 = -c^2 dt^2 + \frac{a^2(t)}{a_0^2} 
           [ a_0^2 d\chi^2 + a_0^2 \chi^2(d\theta^2+\sin^2\theta d\phi^2) ].
\end{equation}
The expansion factor $a(t)$ has as its present value 
$a_0 = R_0/\Delta$, where $R_0 \equiv c/H_0$ and 
$\Delta = (|\Omega_0+\lambda_0-1|)^{1/2}$ 
if the value inside the square-root
is not zero and $\Delta=1$ if it is zero.
Among several relevant cosmological
distances, we use the parametric distance $\chi$, which is
equivalent to the (filled-beam) standard 
angular distance (see Ref. \cite{FFKT}).

Galaxies are modeled as SIS; therefore, 
their lensing properties are
determined by a one-dimensional velocity dispersion, $\sigma$, 
which is related to the galaxy luminosity by 
$ L/L^\ast = (\sigma/\sigma^\ast)^\gamma $
where $\gamma=4$ for E and SO galaxies
and $\gamma=2$ for for S galaxies \cite{Fab76-Tul77,FT}. 
The luminosities of the galaxies follow
the Schechter function
\begin{equation}
   \Phi(L)dL = \Phi^\ast \left(\frac{L}{L^\ast}\right)^{\alpha_g}
               \exp (-L/L^\ast) \frac{dL}{L^\ast}
\end{equation}
where $\alpha_g=-1.1$ \cite{Efs88}.
The exact values of the typical luminosity of a galaxy, $L^\ast$, and
its normalization constant, $\Phi^\ast$, are important
in lensing probability estimates, but are irrelevant in this work
because only the functional shape of the distribution matters.
If there are galaxy evolutions, $\Phi^\ast$ and $L^\ast$ become
functions of $\chi$ (or redshift).
 
Under these assumptions, the lensing probability by one type
of galaxy is simply given as \cite{GPL}
\begin{equation}\label{tauone}
  d^2\tau = \pi a_0^3 (\alpha^\ast)^2 \Phi^\ast
          \left(\frac{L}{L^\ast}\right)^{\alpha_g+4/\gamma}
          e^{-L/L^\ast} 
          \left[\frac{(\chi_S-\chi_L)^2 \chi^2}{\chi_S^2}\right]
          d\left(\frac{L}{L^\ast}\right) d\chi_L 
\end{equation}
where $\chi_S$ and $\chi_L$ are the parametric distance to the
source and the lens, respectively, 
and $\alpha^\ast \equiv 4\pi (\sigma^\ast/c)^2$,
the bending angle for an $L^\ast$ galaxy. Since the image separation
$ \Delta\theta = 2 \alpha^\ast (L/L^\ast)^{2/\gamma} 
  (\chi_S-\chi_L)/\chi_S $,
the probability can be rewritten as a function of $\Delta\theta$ or
$\varphi \equiv \Delta\theta/(2\alpha^\ast)$:
\begin{eqnarray}\label{tautwo}
  d^2\tau & = & \frac{\pi}{2} \gamma a_0^3 (\alpha^\ast)^2 \Phi^\ast 
          \varphi^{(\alpha_g+1)\gamma/2+1} 
          \exp \left[ -\bigl(\frac{\chi_S-\chi_L}{\chi_S}\bigr)^{-\gamma/2}
               \varphi^{\gamma/2} \right] \nonumber\\ 
          & & \times \frac{(\chi_S-\chi_L)^2\chi_L^2}{\chi_S^2}
              \left[\frac{\chi_S-\chi_L}{\chi_S}\right]
                ^{-\gamma(\alpha_g+1+4/\gamma)/2} d\varphi d\chi_L .
\end{eqnarray}
The total probability of lensing by all types of galaxies is
\begin{equation}\label{tautotal}
  \tau = \sum_{i=\hbox{E,S0,S}} 
         \int_{0}^{\infty} d\varphi_i
         \int_{0}^{\chi_S} d\chi_L
         \frac{d^2\tau_i}{d\varphi_i d\chi_L}
\end{equation}
where $d\tau_i$ and the other symbols with subscript $i$ refer to
the respective values for E, S0, and S galaxies.

All lens observations and surveys
suffer from some kind of angular selection bias mainly due to the
finite angular resolution.
If this bias is taken account, 
the probability distribution function (PDF) of observed 
image separations becomes
\begin{eqnarray}
   \frac{dP}{d\Delta\theta} 
         & = & \frac{1}{\tau}
               \int_{0}^{\chi_S} \sum_{i}
               \left(\frac{d^2\tau_i}{d\chi_L d\Delta\theta}\right)
               d\chi_L \\
         & = & \frac{ \sum_{i}(2\alpha^\ast_i)^{-1}\gamma_i
               F^\ast_i [\Gamma(\alpha_g+1+4\gamma_i^{-1})]^{-1}
               \int_{0}^{\chi_S} d\chi_L G_i(\chi_L,\varphi_i) }
                    { \sum_{i}\gamma_i
               F^\ast_i [\Gamma(\alpha_g+1+4\gamma_i^{-1})]^{-1}
               \int_{0}^{\infty} d\varphi_i
               \int_{0}^{\chi_S} d\chi_L G_i(\chi_L,\varphi_i) } 
\end{eqnarray}
with
\begin{eqnarray}
    G_i(\chi_L,\varphi_i) & = &
               f_s (\Delta\theta) 
               \left[ \frac{(\chi_S-\chi_L)\chi_L^2}{\chi_S} \right]
               \left[ \frac{\chi_S}{\chi_S-\chi_L}\varphi_i 
                      \right]^{\gamma_i(\alpha_g+1)/2+1} \nonumber\\
           & & \times \exp \left[ -\Big(
                      \frac{\chi_S}{\chi_S-\chi_L}\Big)^{\gamma_i/2}
                      \varphi_i^{\gamma_i/2} \right]
\end{eqnarray}
where $\Gamma$ is the gamma function$,
f_s(\Delta\theta)$ is a function
describing the angular resolution bias
with $\varphi_i=\Delta\theta/(2\alpha^\ast_i)$, and 
$F^\ast_i \equiv \pi\Phi^\ast_i (\alpha^\ast_i)^2 
 R_0^3 \Gamma(\alpha_g+1+4\gamma_i^{-1})$ is the dimensionless parameter
for the effectiveness of matter in producing double images 
for each type of galaxy \cite{TOG,FT}.
We take the values for the various parameters from Ref. \cite{FFKT}. 
The cumulative distribution function (CDF)
\begin{equation}\label{cdf}
    P(\Delta\theta) = \int_{0}^{\Delta\theta}
                      \left(\frac{dP}{d\Delta\theta'}\right) d\Delta\theta'
\end{equation}
is calculated from $dP/d\Delta\theta$.

It can be proved with ease that in a flat universe the distribution is
independent of $\chi_S$ (or $z_S$, the redshift of the source), regardless
of the value of $\lambda_0$ or the functional form of 
the angular resolution bias \cite{GPL}. 
This implies, regardless of the source distribution
in space, any multiple-image lensing event should have the same
image separation distribution as long as the universe is flat,
which makes various statistical tests very simple.
However, this is true only when
there is no evolution in the lensing galaxies.

\section{Observed Multiple-Image Systems and Angular Selection Bias}
\label{sec:obs}

The image
separation is not affected by the absolute brightness of the 
sources as is the case with the lensing probability.
However, it is affected by the finite angular resolution
and by the dynamic range of the real observations; therefore,
the angular selection bias is a function of
the image separation and the relative brightness of each image.
Hence, to analyze statistically the distribution of image separations
of lens systems discovered in specific surveys, one
should carefully consider both effects as in Ref. \cite{Mao93}.
However, since we approximate the lens by SIS in this work
and since a given SIS lens produces
always the same image separation regardless of the source position,
the angular selection bias will be a function only of the image separation.

Since most of the lens cases are not discovered in one systematic survey
and since each survey or observation is done under different conditions,
a single universal function does not exist which can faithfully
represent the angular selection effect in various different observations.
Hence, we must use different specific angular selection biases for different
observations. In this work, we consider four kinds of selection biases 
suitable for specific surveys: no selection
bias (NO), the one for the HST snapshot survey \cite{Mao93,Koc93a} (HST), 
four similar selection biases appropriate for ground 
optical lens surveys \cite{Koc93a} (PSF1.0, PSF0.7, EYE1.0, EYE0.7)
one for a radio survey \cite{Bur92,Koc93b} (MG-C).
Although the original forms for the selection 
biases HST, PSF1.0, PSF0.7, EYE1.0,
and EYE0.7 depend on the brightness ratio of the images,
we can disregard this dependence for the reasons explained above. 
Hence, we use Kochanek's
completeness function \cite{Koc93a} as the selection bias function
because the image separation statistics are independent of the source
brightness or the redshift distributions.

The lens samples and their corresponding selection biases 
with descriptions are listed in Table \ref{tableSELECTION}
and shown in Fig. \ref{figPDF}: Inverted {\it big triangles},
{\it small triangles}, and {\it circles}
denote the image separations of systems with a single galaxy
lens or unknown lens, systems with a lens other
than a single galaxy, and radio rings.
We use only the accepted cases of multiple-image systems according
to the criterion of Ref. \cite{Sur}. 
Also, we basically exclude systems like 0957 (lensed by
a galaxy and a cluster) and MG2016 (lensed by two galaxies)
because we only consider single-galaxy lensing. However,
the two systems are included in a special case for comparison.
In systems like H1413 and MG0414,
we do not know what kind of object is responsible for lensing. 
However, excluding those systems could introduce
a bias against a small-separation system which has a less massive
galaxy and, therefore, less chance of the lens being detected.
For that reason, we include them as single-galaxy lens systems. 
Although system Q1208 is classified only as a proposed case
in Ref. \cite{Sur}, it is meaningless to 
discuss the HST snapshot survey result
without it \cite{Mao93}, and we include it only for the HST survey.
The test result regarding the HST survey should be taken with some cautions.
Radio rings are included only in MG-C radio survey samples.

\section{Statistical Test} 
\label{sec:stat}

Figure \ref{figPDF} shows the PDF's of the
image separations for each angular selection bias. 
All of them peak around sub-arcsecond separations. 
However, all of the observed single-galaxy
lens systems ({\it big triangles})
have image separations larger than 1\arcsec,
showing a discrepancy with theoretical expectations. 
To access the degree of discrepancy statistically,
we use Monte-Carlo (MC) simulations\cite{Mao93} and the 
Kolmogorov-Smirnov (KS) test.

\subsection{Monte-Carlo Test}

We define a likelihood function \cite{Mao93}
\begin{equation}
    \ell \equiv \ln \left( \prod_{i=1}^{n} \frac{dP}{d\Delta\theta} 
             \big|_{\Delta\theta_i} \right),
\end{equation}
and generate $10^4$ sets of four to six random $\Delta\theta_i$'s
following the calculated PDF
and their respective $\ell$ values. These are compared with
the $\ell$ values for the observed $\Delta\theta_i$'s, and the number 
of MC sets having $\ell$ values less than
that of the observed sample is counted. 
The result is expressed in (absolute) probability in Table \ref{tableMC}. 
We can see that most models are not rejected and
that only the no-selection-bias model is rejected
if we include the large separation, non-single-galaxy lens cases,
MG2016 and Q0957. Also, the model with 
the MG-C selection bias is rejected at 
$\sim 90\%$ level by the MG-C band survey samples.

\subsection{Kolmogorov-Smirnov Test}

The MC simulation is quite versatile, 
being applicable even to the case where the 
distribution differs for each sample data.
However, the test mainly focuses on whether the
data occur at the probable part of the distribution.
It does not test whether the data are clustered or not.
The KS test is complementary to the
MC simulation because it has good discriminative
power on the clustering of the data.
Hence, we use the KS two-sided test to find out
if the calculated distribution is significantly
different from that of the observed samples. This is possible
because in a flat universe the distribution of 
separation angles is the same regardless of the source redshift,
thus making the KS test easily applicable. 

The CDF's constructed from the
observed lens samples ({\it jagged lines}) and from 
the calculated distribution ({\it smooth curves})
are shown in Fig. \ref{figCDF}:
(a) the {\it dot-dashed} curve is CDF for no selection bias (NO),
the {\it solid} line CDF from lens systems O-ALL,
the {\it dotted} line from O-SINGLE,
and the {\it dashed} line from O/R-SINGLE;
(b) the {\it dotted} curve for bias HST
and the {\it solid} line from corresponding lens systems;
(c) the {\it dotted} curve for bias PSF1.0,
the {\it short-dashed} curve for bias PSF0.7,
the {\it long-dashed} curve for bias EYE1.0,
the {\it dot-dashed} curve for bias EYE0.7
and the {\it solid} line from corresponding lens systems;
(d) the {\it dotted} curve for bias MG-C
and the {\it solid} line from corresponding lens systems.
The comparisons between the observed and the calculated CDF's show
the largest differences for the NO bias and the MG-C bias models.

The results of the KS test applied to each bias case are shown
in Table \ref{tableKS}. The numbers show the confidence
of rejection, and only NO-O-ALL and MG-C are rejected
with more than 95\% confidence. This is in agreement with
the MC test results. Hence, we can say that {\sl although the
observed optical lens systems show somewhat different
distribution from theoretical ones, the difference is not statistically
significant}. Only the radio lens systems show 
statistically significant inconsistencies with the ``standard''
lens statistics calculation. However, this is based on only four systems.
If we dismiss lens system MG2016 which has two lensing galaxies,
the sample consists of only three systems.
Hence, although a statistically significant conclusion
can be drawn formally, it is based on a few lens systems and should
be taken with cautions.

\subsection{Without the $(3/2)^{1/2}$ Factor}

Turner et al. \cite{TOG} were the first to 
introduce the correction factor $(3/2)^{1/2}$ to
convert the observed velocity dispersion (of
luminous matter) into that of dark matter for
E and S0 galaxies (also see Ref. \cite{FT}). However, Kochanek
argues against this correction \cite{Koc93b}
and estimates a ``90\% confidence range'' 
in the velocity dispersion which
is significantly smaller than the $(3/2)^{1/2}$ corrected value.
Hence, we repeat the calculation and the test without
the $(3/2)^{1/2}$ factor. 
Since the image separation in a SIS lens is directly
proportional to the square of the velocity dispersion,
the mean image separation get smaller by 67\% and
the whole distribution moves to smaller values.
The PDF's with and without the $(3/2)^{1/2}$ factor are 
compared in Fig. \ref{fig3_2} ({\it solid} for
with and {\it dotted} for without).
The result of the MC test
is summarized in Table \ref{tableMC}: All models, except
HST and EYE1.0, have probabilities less than 10\%, and the 
no-selection-bias models and MG-C model have probabilities less than 5\%. 
The result of the KS test is shown in Table \ref{tableKS}:
Again, the confidence of rejection is $\gtrsim 90\%$ for
all models except HST, and no-selection-bias models
and the MG-C model are rejected by the observed lens systems with
more than 97\% confidence.
To summarize, if the $(3/2)^{1/2}$ factor is really unnecessary,
the observed lens data reject the ``standard'' lens statistics
models, marginally or strongly depending on the angular
selection biases and the corresponding data sets used.

\section{Discussion}
\label{sec:disc}

There are several factors which can contribute
to the uncertainties in the probability distribution
and in the statistical test.
Firstly, the test depends sensitively on the observation
sample due to the small number of appropriate lensing cases.
From the analysis, we have seen that the
conclusion does depend on which observed lens systems are chosen.
This problem will disappear when we have enough lens systems.

We may increase the observed sample size by including
other lens systems like Q0957+561 where the lensing galaxy is aided by
a cluster. However, then the theoretical calculation should be modified
accordingly to incorporate such multiple deflector cases.
We think it is misleading to include lens system like Q0957
in the observed sample and to compare it with the theoretical
distribution calculated by assuming only one galaxy 
as the deflector. However, this was done in many previous works
\cite{FFKT,Mao93,Koc93a,Koc93b}.

It is also possible to use the measured
redshifts of the deflectors and to calculate the distribution
of image separations for a given lens redshift and source
redshift. This approach would be useful if we had enough lens
systems because the distribution in this case would depend
on $\Omega_0$ and $\lambda_0$ individually and would be relatively
insensitive to magnification bias. However, the number of lens
systems found so far is too small to make any statistically
significant discrimination \cite{Lee94}.
We need more clean lens cases to make the test meaningful.

Secondly, the presence of cores in the galaxies and asphericity
certainly affects the distribution. The core radii of galaxies are
generally small enough to make SIS a good approximation
\cite{TOG,Mao93}. If the core radii are not smaller than
the typical image separations in the lens plane, the expected
separation of images will be smaller and the discrepancy
will be larger.
The ellipticity can affect lensing frequencies
via the magnification bias \cite{Wal93,Koc93b}. 
However, the distribution of image separations
is relatively insensitive to the magnification bias---
the magnification bias can affect the distribution only indirectly
through the angular selection bias which is weakly linked
to the magnification bias \cite{Koc93a}. It is possible
to have high magnification events having much smaller 
separations than the characteristic values of the SIS cases.
However, these types of events are expected to be rare \cite{FT}.

The rather strong disagreement between the observation
and the theoretical expectation, 
when we omit the $(3/2)^{1/2}$ factor, 
is not apparent in the extensive maximum
likelihood study of Kochanek \cite{Koc93b}. The main reason
for this difference is the observed lens systems used.
Lens systems Q1208+101, B1938+666, and B0218+356,
all having $\lesssim 1\arcsec$ separations, are
included in Kochanek's sample---
they are omitted from our sample 
because they do not pass the criterion for the accepted
case or because the image separation is not known yet.

Now, we can think of possible reasons for the discrepancy
between the probability distribution from the theory and
that from the observational data.
The main ingredients of the ``standard'' model can be 
divided into two parts: one regarding cosmology and the
other regarding galaxies. To determine if different cosmological
model would change the test result, we tried two non-flat
cosmological models (the open model with $\Omega_0=0.1$, 
$\lambda_0=0$ and the closed model with $\Omega_0=2$, $\lambda_0=0$).
We applied a MC simulation test to these models.
The probabilities from the MC simulations increased by $10\sim30\%$
for the closed universe and decreased by $20\sim30\%$ for the open
universe because the image separation is smaller
in an open universe and larger in a closed universe than in a
flat universe \cite{GPL}. Yet, the result was inconclusive again.
Hence, the cause of the inconsistency between the ``standard'' model
and the observation is likely to be in our poor understanding
of galaxies rather than in the cosmological models. If this is
the case, all previous works using the ``standard'' lensing
statistics model to probe the cosmological parameters should 
be taken with caution.

\section{Summary}
\label{sec:sum}

We calculated the distribution of image separations in the ``standard''
gravitational lens statistics model where the universe is flat and
lensing galaxies are modeled by SIS's which
follow the Schechter luminosity function and whose
comoving density is kept constant. 
The calculated distribution of image separations, 
incorporating the angular selection
bias, is compared, through the MC and the KS tests, 
with that seen in the observed lens systems.

The calculated distribution implies that a significant fraction of
observed single-galaxy lens systems should have image separations
$\lesssim 1\arcsec$, whereas most of the observed lens systems
have separations larger than $1\arcsec$. However, the distribution
is wide enough that the observations do not reject
the ``standard'' lensing statistics model with enough
statistical significance. The model without consideration
of the angular selection bias is nearly ruled out, and the radio
data strongly rules out the ``standard'' model. However, this conclusion,
although statistically significant,
is based on only four or three lens cases. 

If the correction factor $(3/2)^{1/2}$---for converting
the observed velocity dispersion to that of dark matter---
is not used in the lens statistics calculations, the
``standard'' model is mostly rejected by the observations,
although the ground-based survey is marginally compatible
with the ``standard'' model.
Also, larger core radii produce smaller
image separations, and the discrepancy between the ``standard''
model and the observation will be greater,
though the core radii of most galaxies are
thought to be small enough to make SIS a good approximation.
	
The greatest difficulty in this or similar work is the 
small number of observed lens systems suitable for this kind
of statistical analysis. We need to have more
``clean'' lens cases, meaning a single galaxy acting as lens.
We hope that projects like Sloan Digital Sky Survey will answer most
of the problems we have now.

\acknowledgments
We thank the referee for useful comments.
This work was supported in part by Korea Research Foundation and
in part by Korean Science and Engineering Foundation.

\begin{figure}
  \caption{Probability distribution functions of image separations
           in multiple-image lens systems for various angular resolution
           biases: NO ({\it solid line}), HST ({\it dotted line}),
           PSF1.0 ({\it short dashed line}), 
           PSF0.7 ({\it long dashed line}),
           EYE1.0 ({\it short dash-dotted line}),
           EYE0.7 ({\it long dash-dotted line}), and
           MG-C ({\it long dash-short dashed line}). Larger triangles
           mark the image separations in the lens systems 
           PG1115, Q2237, Q0142, H1413, MG0414, and B1422.
           Smaller triangles mark those in Q1208 and MG2016.}
\label{figPDF}
\end{figure}

\begin{figure}
  \caption{Cumulative distribution functions (CDF's)---fraction of cases with
           separations smaller than $\Delta\theta$---
           for various resolution biases and those from observed
           lens cases: 
           (a) the {\it dot-dashed} curve is CDF for no selection bias (NO),
           the {\it solid} line CDF from lens systems O-ALL,
           the {\it dotted} line from O-SINGLE,
           and the {\it dashed} line from O/R-SINGLE;
           (b) the {\it dotted} curve for bias HST
           and the {\it solid} line from corresponding lens systems;
           (c) the {\it dotted} curve for bias PSF1.0,
           the {\it short-dashed} curve for bias PSF0.7,
           the {\it long-dashed} curve for bias EYE1.0,
           the {\it dot-dashed} curve for bias EYE0.7,
           and the {\it solid} line from corresponding lens systems;
           (d) the {\it dotted} curve for bias MG-C
           and the {\it solid} line from corresponding lens systems.}
\label{figCDF}
\end{figure}

\begin{figure}
  \caption{PDF's with [{\it solid}] and without [{\it dotted}]
           the $(3/2)^{1/2}$ correction factor for no selection bias (a)
           and for MG-C selection bias (b).}
\label{fig3_2}
\end{figure}


\begin{table}
\caption{Relevant multiple image gravitational lens systems.}
\begin{tabular}{lll}
       Lens system 
     & $\Delta\theta$\tablenotemark[1]
     & $z_S$ \\
\tableline
	PG1115\tablenotemark[2]		& 2.3 & 1.72 \\
	Q2237\tablenotemark[2]		& 1.8 &	1.69 \\
	Q0142\tablenotemark[2]		& 2.2 & 2.72 \\
	H1413\tablenotemark[2]		& 1.1 & 2.55 \\
	MG2016\tablenotemark[3]		& 3.8 & 3.27 \\
	Q0957\tablenotemark[4]		& 6.1 & 1.41 \\
	MG0414\tablenotemark[5]		& 3.0 & 2.64 \\
	B1422\tablenotemark[5]		& 1.3 & 3.62 \\
	Q1208\tablenotemark[6]		& .47 & 3.80 \\
	MG1131\tablenotemark[7]		& 2.1 & 1.13? \\
	MG1654\tablenotemark[7]		& 2.1 & 1.74 \\
\end{tabular}
\tablenotetext[1]{Maximum image separation.}
\tablenotetext[2]{Discovered in an optical observation. 
                  Lensed by a single galaxy.}
\tablenotetext[3]{Discovered in an optical observation. 
                  Lensed by two galaxies.}
\tablenotetext[4]{Discovered in an optical observation. 
                  Lensed by a galaxy plus cluster.}
\tablenotetext[5]{Discovered in a radio observation. 
                  Lensed by a galaxy.}
\tablenotetext[6]{Classified as a proposed case\cite{Sur}.}
\tablenotetext[7]{Radio ring.}
\label{tableSAMPLE}
\end{table}

\begin{table}
\caption{Angular selection bias and corresponding lens systems.}
\begin{tabular}{ll}
       Selection Bias
     & Lens Systems \\
\tableline
        NO\tablenotemark[1]	
      & 1115, 2237, 0142, 1413, 2016, 0957 (O-ALL\tablenotemark[2])\\	
        NO\tablenotemark[1]
      & 1115, 2237, 0142, 1413  (O-SINGLE\tablenotemark[3]) \\	
        NO\tablenotemark[1]
      & 1115, 2237, 0142, 1413, 0414, 1422 (O/R-SINGLE\tablenotemark[4]) \\
        HST\tablenotemark[5] 
      & 1115, 2237, 0142, 1413, 1208      \\
        PSF1.0\tablenotemark[6]
      & 1115, 2237, 0142, 1413           \\
        PSF0.7\tablenotemark[7]
      & 1115, 2237, 0142, 1413           \\
        EYE1.0\tablenotemark[8]
      & 1115, 2237, 0142, 1413           \\
        PSF0.7\tablenotemark[9]
      & 1115, 2237, 0142, 1413           \\
        MG-C\tablenotemark[10]
      & 2016, 0414, 1131, 1654           \\
\end{tabular}
\tablenotetext[1]{No selection bias.}
\tablenotetext[2]{Lens systems discovered in optical observations.}
\tablenotetext[3]{Lens systems discovered in optical observations. 
                  Lensed by a single galaxy or lens not known.}
\tablenotetext[4]{Lens systems discovered in optical or radio observations. 
                  Lensed by a single galaxy or lens not known.}
\tablenotetext[5]{Selection bias for the HST snapshot survey.}
\tablenotetext[6]{Selection bias for ground-based optical observations
                  with the PSF subtraction method 
                  and a seeing FWHM=1.0'' \cite{Koc93a}.}
\tablenotetext[7]{Selection bias for ground-based optical observations
                  with the PSF subtraction method 
                  and a seeing FWHM=0.7'' \cite{Koc93a}.}
\tablenotetext[8]{Selection bias for ground-based optical observations
                  with visual examination
                  and a seeing FWHM=1.0'' \cite{Koc93a}.}
\tablenotetext[9]{Selection bias for ground-based optical observations
                  with visual examination
                  and a seeing FWHM=0.7'' \cite{Koc93a}.}
\tablenotetext[10]{MG C band survey \cite{Bur92,Koc93b}}
\label{tableSELECTION}
\end{table}

\begin{table}
\caption{Monte-Carlo simulation results.} 
\begin{tabular}{lcc}
     Selection Bias/Samples & 
     \multicolumn{2}{c}{Probability\tablenotemark[1]} \\
     & With $(3/2)^{1/2}$ & Without $(3/2)^{1/2}$ \\ \tableline
     NO-O-ALL			& .01	& .00	\\
     NO-O-SINGLE		& .37	& .04	\\
     NO-O/R-SINGLE		& .30	& .01	\\
     HST			& .56	& .11	\\
     PSF1.0			& .47	& .08	\\
     PSF0.7			& .42	& .06	\\
     EYE1.0			& .52	& .12	\\
     EYE0.7			& .44	& .07	\\
     MG-C			& .09	& .00	\\
\end{tabular}
\tablenotetext[1]{Fraction of MC draws having likelihood
                  smaller than that of the observed data.}
\label{tableMC}
\end{table}

\begin{table}
\caption{Kolmogorov-Smirnov test results.} 
\begin{tabular}{lcc}
     Selection Bias/Samples & 
     \multicolumn{2}{c}{Confidence of Rejection\tablenotemark[1]} \\
     & With $(3/2)^{1/2}$ & Without $(3/2)^{1/2}$ \\ \tableline
     NO-O-ALL			& .94	& .998	\\
     NO-O-SINGLE		& .71	& .97	\\
     NO-O/R-SINGLE		& .87	& .995	\\
     HST			& .17	& .81	\\
     PSF1.0			& .53	& .92	\\
     PSF0.7			& .60	& .94	\\
     EYE1.0			& .39	& .89	\\
     EYE0.7			& .50	& .92	\\
     MG-C			& .98	& .999	\\
\end{tabular}
\tablenotetext[1]{Confidence to reject the null hypothesis
                  that data come from the model distribution.}
\label{tableKS}
\end{table}

\widetext

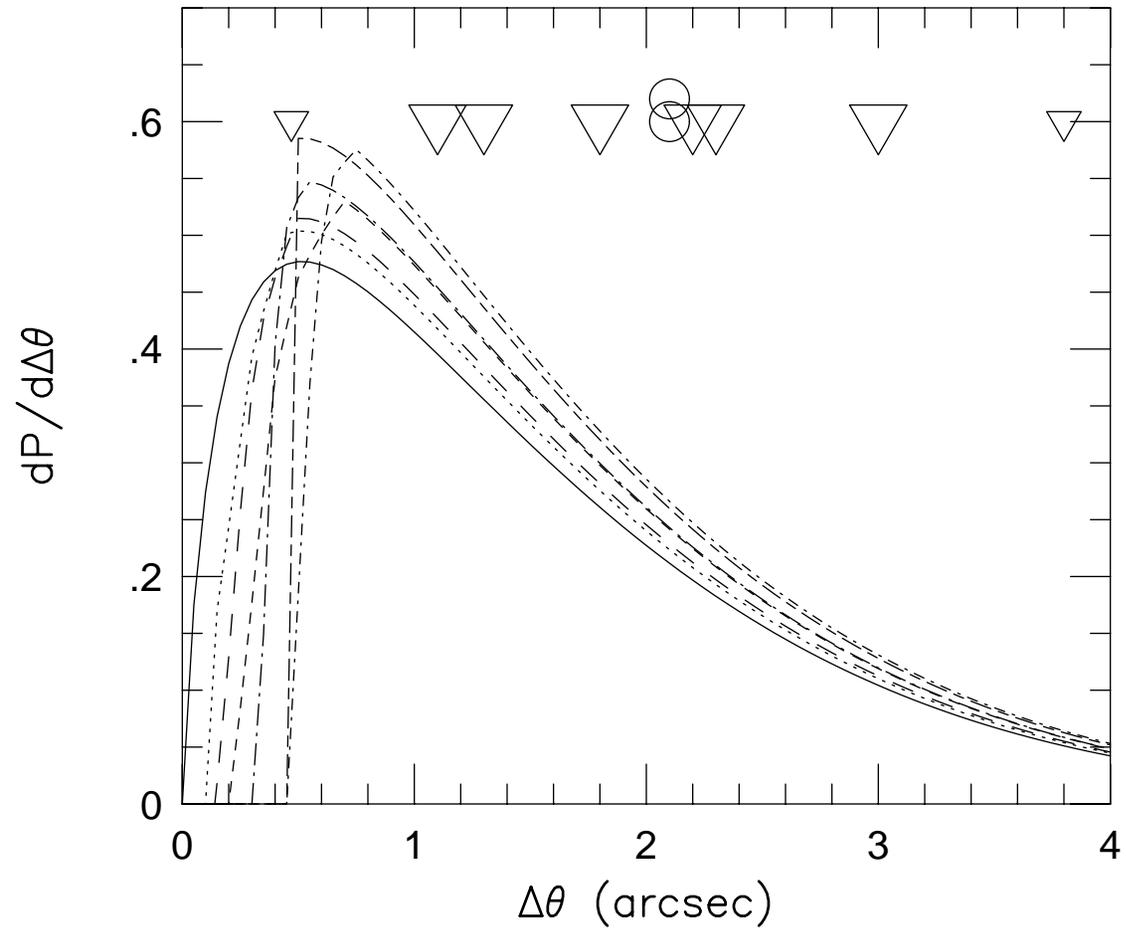
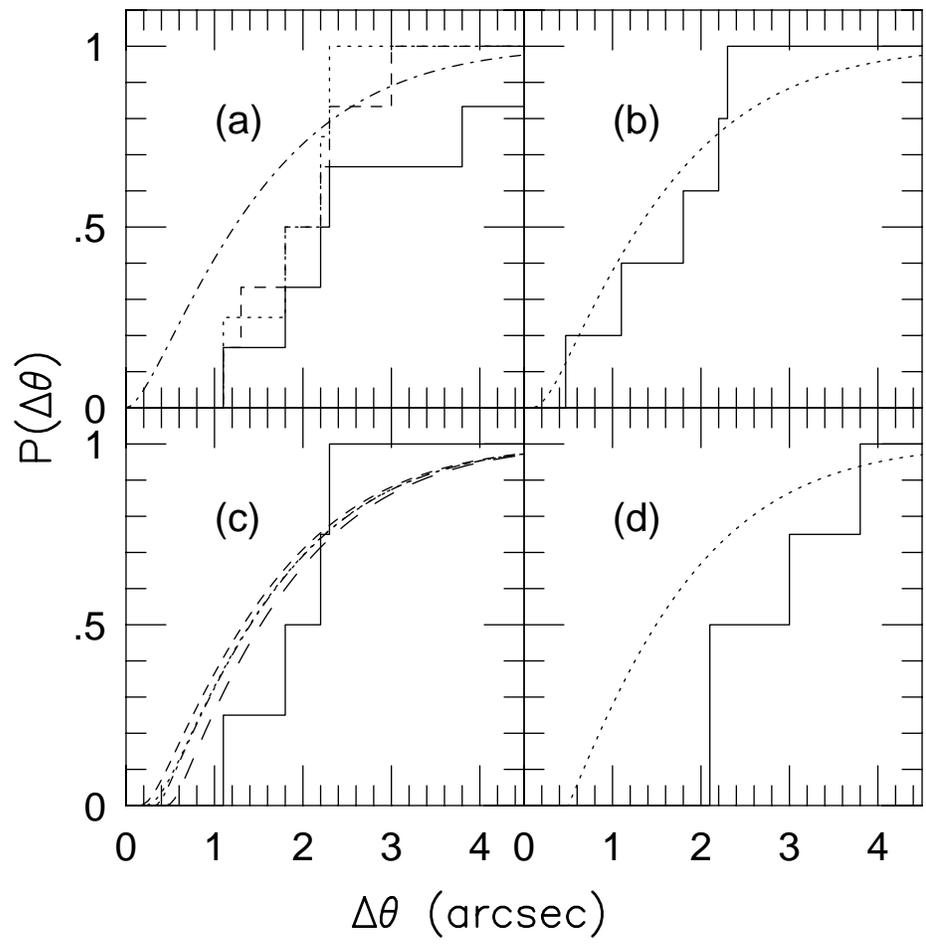
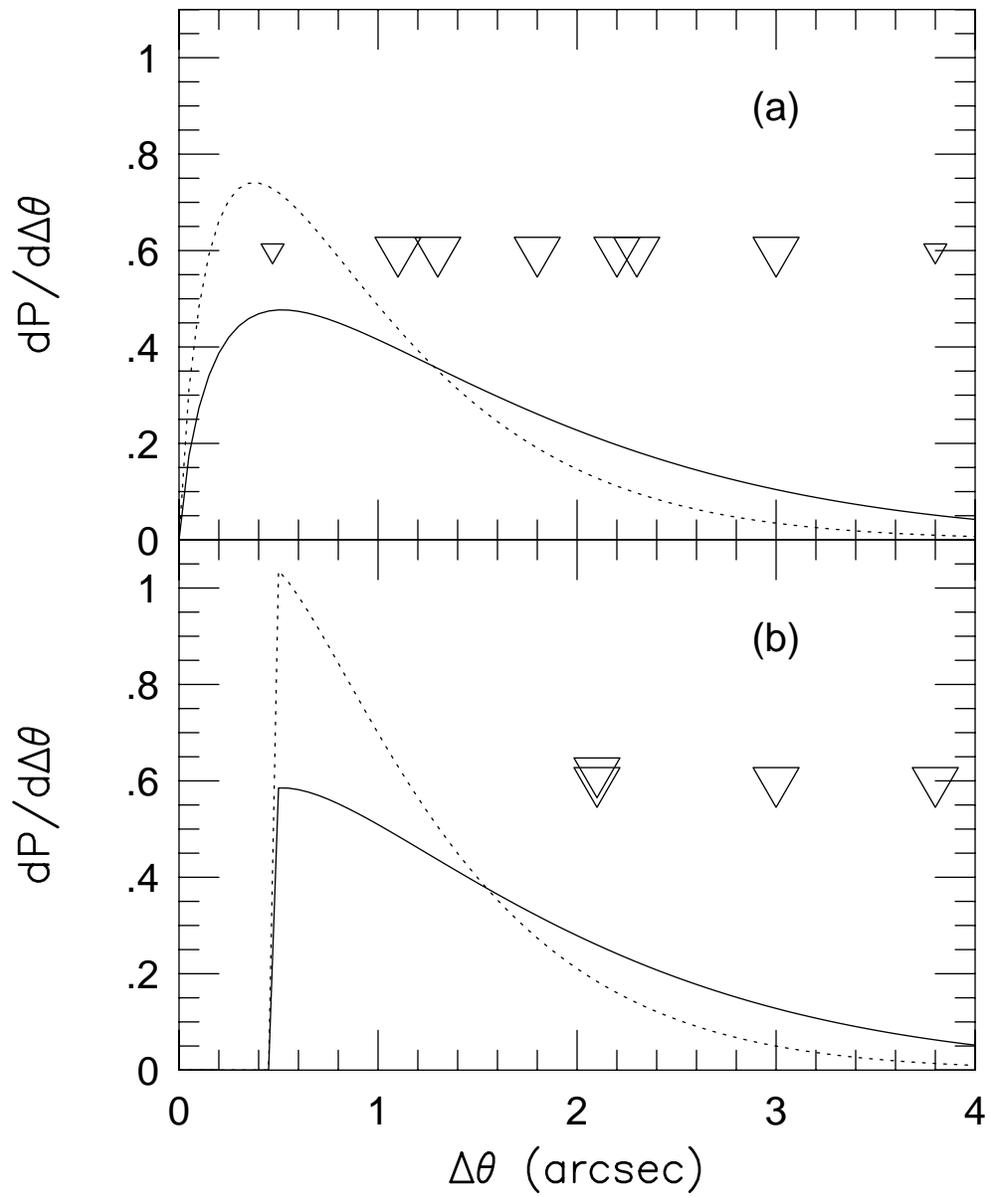
\end{document}